\documentclass[10pt, a4paper, conference]{IEEEtran}
\usepackage{cite}

\ifCLASSINFOpdf
  \usepackage[pdftex]{graphicx}
\else
  \usepackage[dvips]{graphicx}
\fi
\usepackage{url}

\usepackage{listings}
\usepackage{color}
\usepackage{subcaption}
\usepackage{booktabs}
\usepackage{amssymb}
\usepackage{amsmath}
\usepackage{todonotes}

\definecolor{mygreen}{rgb}{0,0.6,0}
\definecolor{mygray}{rgb}{0.5,0.5,0.5}
\definecolor{mymauve}{rgb}{0.58,0,0.82}
\definecolor{light-gray}{gray}{0.95}

\usepackage{xargs}
\newcommandx{\unsure}[2][1=]{}
\newcommandx{\change}[2][1=]{}
\newcommandx{\info}[2][1=]{}
\newcommandx{\improvement}[2][1=]{}
\newcommandx{\thiswillnotshow}[2][1=]{}

\begin{document}
%
\title{Understand Your Chains: Towards Performance Profile-based Network Service Management}

\author{\IEEEauthorblockN{Manuel Peuster}
\IEEEauthorblockA{Paderborn University\\
manuel.peuster@uni-paderborn.de}
\and
\IEEEauthorblockN{Holger Karl}
\IEEEauthorblockA{Paderborn University\\
holger.karl@uni-paderborn.de}
}


%


\maketitle

\begin{abstract}


Allocating resources to virtualized network functions and services to meet service level agreements is a challenging task for NFV management and orchestration systems. This becomes even more challenging when agile development methodologies, like DevOps, are applied. In such scenarios, management and orchestration systems are continuously facing new versions of functions and services which makes it hard to decide how much resources have to be allocated to them to provide the expected service performance. 
One solution for this problem is to support resource allocation decisions with performance behavior information obtained by profiling techniques applied to such network functions and services.


In this position paper, we analyze and discuss the components needed to generate such performance behavior information within the NFV DevOps workflow. We also outline research questions that identify open issues and missing pieces for a fully integrated NFV profiling solution. Further, we introduce a novel profiling mechanism that is able to profile virtualized network functions and entire network service chains under different resource constraints before they are deployed on production infrastructure.

\end{abstract}



%
\IEEEpeerreviewmaketitle

\section{Introduction}
\label{sec:intro}

Network softwarization is seen as one of the key concepts to cope with the high agility required by the upcoming 5th generation of networks. 
One of its benefits is the possibility to apply software engineering concepts, like DevOps, to the development cycle of virtualized network functions (VNF) which results in reduced turnaround times, faster time-to-market, and reduced operational expenses (OPEX).

\begin{figure}[ht]
	\centering
	\includegraphics[width=0.8\columnwidth]{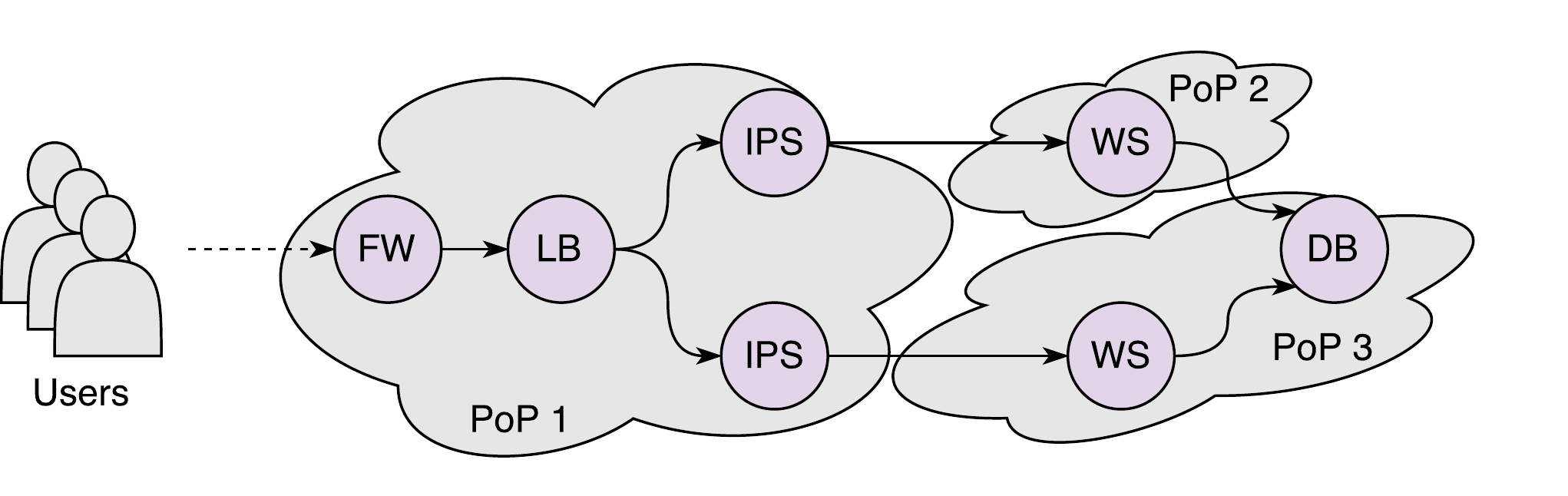}
	\caption{Network service chain deployed on multiple PoPs consisting of firewall (FW), load balancer (LB), intrusion preventions system (IPS), web server (WS), and database  (DB)}
	\label{fig:ns}
\end{figure}

Even though the softwarization of network functions comes with many benefits it also introduces new  challenges and problems. One of them is the enforcement of service level agreements (SLA)  in such dynamic software environments. This question is aggravated by VNFs being deployed as part of complex network service chains (NS), containing multiple VNFs (Fig.~\ref{fig:ns}). These chains are often deployed between the end users and services and are distributed across multiple points of presence (PoPs). As a result, enforcing the quality of service (QoS) and especially the quality of experience (QoE) of the entire chain is crucial to meet user expectations. 

Existing approaches for these problems rely on live monitoring solutions. In such systems, performance data is continuously collected and the deployment of the network service is adapted to meet the SLAs. However, this has the downside that it is not possible to make statements about the expected performance and resource requirements of a network service prior to its deployment. It also makes the consequences of an adaption decision difficult to foresee since the management system has no concrete knowledge about the service behavior under adapted resource allocations. Consequently, the NFV and research community is looking for benchmarking and profiling solutions that produce performance profiles that give insights about the expected service performance for given resource allocations~\cite{rfc.draft}\cite{rosa2015vbaas}.

In this position paper, we analyze and discuss the missing components to gather, process, and use additional profiling-based performance information to improve the NFV development cycle and optimize service deployments. In particular, we focus on the DevOps concept and its interactions between service development and operation, in Section~\ref{sec:nfvprofiling}. In Section~\ref{sec:questions}, we identify and formulate a list of research questions that need to be answered to provide a fully automated profiling solution that not only supports single network functions but also complex network service chains and can be applied prior to service deployment. These questions also motivate novel management and orchestration (MANO) solutions that utilize performance profiles to improve their resource allocation mechanisms, e.g., scaling decisions.

In the second part of this paper, we introduce our novel profiling mechanism that is able to profile entire service chains on a developer's laptop and generate fine-grained performance profiles (Section~\ref{sec:approach}). After presenting early experimental results that indicate the feasibility of our solution in Section~\ref{sec:measurements}, we discuss existing work in Section~\ref{sec:rw}. Finally, we conclude this paper, which aims to be a first step towards a fully integrated profiling solution for NFV.

\section{Profiling as Part of the NFV DevOps Cycle}
\label{sec:nfvprofiling}

The overall goal of the DevOps methodology is to bridge the gap between development and operation of services~\cite{Kim2015}. New service versions are directly deployed into production after they have been quickly tested in an automated fashion. 

As a result, extensive tests on lab testbeds should be removed from the development cycle. This becomes challenging for NFV where our services are always expected to meet certain SLAs. 
On one hand, it becomes hard for service developers to validate that their changes result in the expected performance improvements before they put their service in production. 
On the other hand, MANO systems will be continuously faced with the management of new service versions, which means that their resource allocation algorithms, e.g., scaling algorithms, have to be continuously adapted. This can be tricky because historical monitoring information, available from old service versions, might not provide correct assumptions about the new version. 

To overcome this, a mechanism is needed that automatically gathers performance information about a service prior to its deployment without requiring dedicated testbeds or other special hardware setups. We call this \emph{offline profiling}.

Another important point that motivates the need for \emph{offline profiling} is based on the assumption that low-level metrics, like throughput, are often not sufficient to perform good resource allocation decisions. Especially for QoE optimizations application-level metrics, like frames/s or lag ratio of a video stream, are more interesting. However, due to encryption and privacy issues, it is not always possible to collect such metrics from operating services, e.g., no deep packet inspection mechanisms (DPI) are available. In an offline profiling solution, in contrast, a developer is allowed to collect all performance metrics he is interested in. It is, for example, possible to add additional measurement VNFs, called \emph{probes}, to the profiled service chain.

\begin{figure}[ht]
	\centering
	\includegraphics[width=1.0\columnwidth]{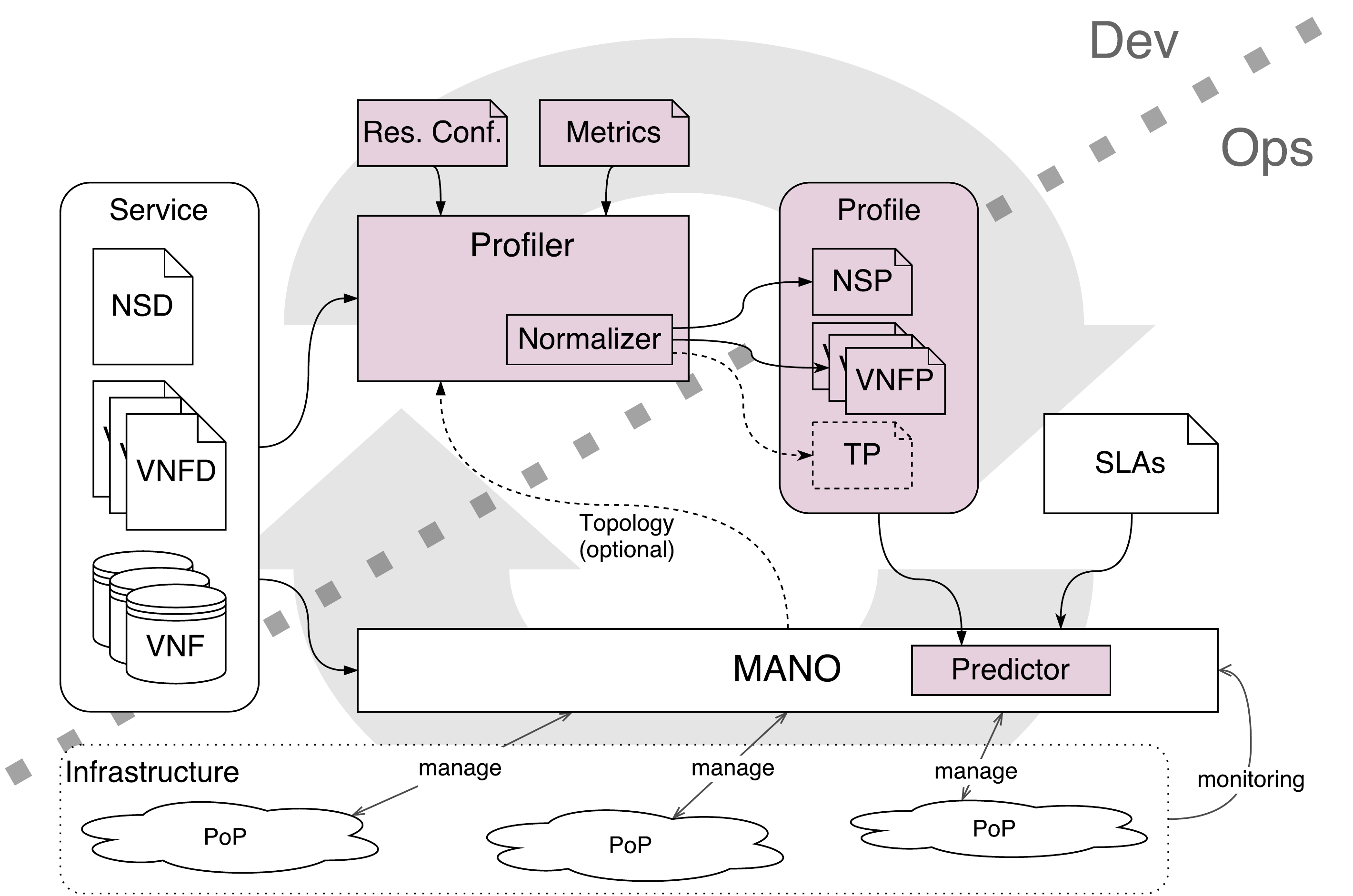}
	\caption{High-level DevOps architecture with \emph{offline profiling}}
	\label{fig:motivation}
\end{figure}

Fig.~\ref{fig:motivation} shows a high-level NFV DevOps architecture. It contains artifacts and components that exist in most of today's architectures including service definitions, consisting of network service descriptors (NSD), VNF descriptors (VNFD), and VNF images as well as the MANO system that manages the service.

We add some components (filled boxes) to this architecture to integrate an \emph{offline profiling} solution into the DevOps cycle. First, there is the main \emph{profiler} component that is part of the service development toolchain and can be executed on the developer's laptop. This profiler gets the service definition and VNFs that should be profiled. Additionally, the developer specifies which resource configurations should be tested throughout the profiling runs and which performance metrics should be collected. The profiler then executes the service and its VNFs with different resource configurations and outputs profiling results for both the network service (NSP) as a whole and each constituting VNF (VNFP). Optionally, topology information about possible target environments can be fed to the profiler. Based on this information special profiling runs can be performed in which the target topology, e.g., a multi-PoP topology with realistic inter-PoP delay, is emulated and the service is tested in this topology. The additional profiling results are called topology profiles~(TP).

The profiling output of such an \emph{offline profiler} highly depends on the host machine on which the profiling run was performed. This makes it hard to reuse them in other environments or compare them. Hence, we foresee a \emph{normalizer} component to be part of the profiling tool that normalizes the results with respect to the underlying machine~\cite{Fleming:1986}.

The final, normalized profiles are then bundled with the service description and pushed to the operation side, namely the MANO system. By doing so, the MANO system has much more information about resource requirements available than in existing approaches. We foresee a \emph{predictor} component inside the MANO system that uses these information to calculate the absolute resources required to meet the given SLAs in the target environment. Since the described performance profiles only provide \emph{relative} profiling data and no \emph{absolute} performance numbers for the target environment, the predictor is required to interpret the available information and predict the resource requirements of the target environment, e.g, by identifying trends in the profiles. 

This approach will still be combined with monitoring-based management solutions and should not be seen as a replacement for performance monitoring functionalities. In the proposed scenario, monitoring information that becomes available after the initial deployment of a service will be used to continuously improve and refine the initial performance profiles and thus the decisions made by the management system. 


\section{Challenges and Research Questions}
\label{sec:questions}

The presented DevOps architecture with \emph{offline profiling} capabilities raises several research questions about the design and functionalities of the new components and interfaces. 

\begin{itemize}
	\item[Q1:] \emph{How to automatically generate detailed performance profiles of network functions prior to their deployment?}
Network functions are complex, multi-layer software systems and modeling their performance or scaling behavior manually or even automatically is not feasible. To this end, a profiling mechanism that executes the function and analyzes its performance as well as used resources is needed.

	\item[Q2:] \emph{How to extend profiling approaches and models to complex service chains?}
Profiling single network functions will already increase the knowledge about the runtime behavior of a service composed of these functions. But such function-specific profiles will not be able to capture all runtime dynamics of complex service chains. To identify bottlenecks and relationships between resources assigned to different parts of the service chain it should be possible to profile the entire chain at once. 
	
	\item[Q3:] \emph{Can locally obtained profiling results be used to predict service behavior in production environments? How good are these predictions?}
Performance numbers obtained by profiling solutions that execute functions and services offline, not in their final production environments, will not provide absolute performance values. Thus, translation and prediction models are needed that identify trends in offline profiling results and use these together with information about the target environment to predict the resulting service performance or resource requirements to meet given SLAs. 
	
	\item[Q4:] \emph{How to include application-level metrics into the profiling process and which benefits can be provided by such application-level profiles?}

Low-level performance metrics, like throughput or delay, are not always well suited to capture the QoE experienced by end-users. Creating performance profiles based on application-level metrics should change this. 

	
	\item[Q5:] \emph{How will next-generation management and orchestration systems utilize performance profiles?}

Existing MANO systems base most of their resource allocation, scaling, and placement decisions on monitoring data collected at runtime. Having detailed performance profiles of a service available as input to these systems before the service is deployed will help to optimize their decisions. Interfaces and description models to annotate services with such profiles are still missing and architectural changes to the control loops implemented in existing systems might be needed.

	\item[Q6:] \emph{How can offline performance profiles be updated and enhanced by online measurements, once a service is deployed and in operation?}
	
It is clear that running services will still be monitored by MANO systems even if offline profiling information is available. These online measurements should be combined with the knowledge from offline profiles to gain the best possible insights into a service's runtime behavior.

	\item[Q7:] \emph{In virtualized environments physical resources are shared between services and can lead to performance interference effects. Can those effects be considered and modeled in an offline profiling approach? Which benefits are provided by the resulting profiles?}

Multiple competing services running on the same physical system will mutually affect their performance. A profiling solution might consider such effects by emulating other services that compete for the same set of resources. 

\end{itemize}

\section{Novel Profiling Approach}
\label{sec:approach}

Our initial work on the topics and questions described in the first part of this paper focuses on the \emph{profiler} component shown in Fig.~\ref{fig:motivation}. In particular, we tried to find a profiling solution that does not rely on expensive, inflexible cloud testbeds and integrates seamlessly into a NFV DevOps cycle.

\subsection{Requirements}

Our profiling solution has to fulfill the following requirements:

\begin{itemize}
	\item[R1:] \emph{Profile production-ready VNFs.} In a DevOps approach time matters. Thus, a profiling system has to be able to execute the same VNFs that will later be executed in the production environment.
	\item[R2:] \emph{Profiling has to be done offline.} Network service and function developers want to quickly check the impact of their changes \emph{before} a service is put to production.
	\item[R3:] \emph{Support profiling of complex service chains.} Profiling a service chain as a whole will give more detailed insights about relative resource requirements of its parts. 
	\item[R4:] \emph{The profiling process has to be fully automated.} NFV is about automation.  Thus the profiling step has to be automated as well.
	\item[R5:] \emph{Profiles should contain fine-grained performance results.}
	Fine-grained performance profiles will better support prediction and decisions algorithms of a MANO system.
\end{itemize}

\subsection{Background: MeDICINE Platform}

Our profiler prototype builds upon an existing emulation platform called \emph{MeDICINE}~\cite{Peuster2016Medicine}. This platform was developed by us and uses a Mininet~\cite{lantz2010network} extension called Containernet~\cite{Containernet.webpage} to execute arbitrary VNFs given as Docker containers~\cite{Docker.webpage} in user-defined, emulated network topologies. This allows the execution of production-ready VNFs as long as they are given as pre-configured Docker containers (R1). The \emph{MeDICINE} platform can be executed locally on a single machine, e.g., the developer's laptop (R2). It also provides built-in solutions to run complex network service chains in (virtually) isolated PoPs (R3) and it is fully scriptable (R4). Besides these functionalities, the \emph{MeDICINE} platform offers an API to control the resources available to each running container, e.g., by using the bandwidth limitation feature of Linux' completely fair scheduler (CFS)~\cite{turner2010cpu} or by assigning a container to a specific set of CPU cores or limit its memory. These fine-grained resource control mechanisms allow us to extend and use the platform for performance profiling tasks (R5) as described in the following.

\subsection{Resource Limit-based Offline Profiling}

There are two types of profiling approaches that can be applied to NFV use cases. The first one utilizes cloud testbeds to execute VNFs in realistic environments under different resource configurations. To do so, a VNF is executed as a VM with a pre-defined resource configuration, e.g., 2 vCPU cores and 2 GB memory, and its performance is measured, e.g., its throughput. After this, the VM is destroyed and a new one with another resource configuration is started, e.g, 4 vCPU core and 2 GB memory. Based on this, performance values for different resource configurations can be measured, which creates a mapping from available resources to resulting performance. This approach provides only a limited set of possible resource configurations (CPU cores, memory, disk space) and it requires a lot of effort to configure and provision the needed VMs.

The second approach executes a single VNF and sends varying amounts of workload to it. During this, its resource consumption, like CPU and memory, is \emph{measured} so that the results reflect a mapping from workload to resource usage. The benefit of this approach is that it comes with less configuration overhead. However, it does not generate results about the behavior of a VNF under different resource limitations, e.g., different numbers of available CPU cores.

In this work, we introduce another solution which combines concepts of both existing approaches; we call it \emph{resource limit-based offline profiling}. The main idea of our profiling approach is to utilize the resource limitation functionalities of container solutions, like Docker, to run VNFs with different resource limitations and measure their performance under these configurations. This allows us to have very precise control of resources available to the tested VNFs. In particular, we can not only control how many cores\footnote{Maximum number of cores is limited by the number of physical cores available in the host machine.} are assigned to the test container but also how much of the available CPU time is used by it. In addition, the container technology also allows us to limit the available memory, swap memory, and block device I/O bandwidth. Optionally, we can profile complex service chains in different network topologies using artificial network delays configured through Mininet's topology API.

It is clear that this profiling solution will not provide absolute performance numbers that can directly be used to calculate the performance of a service in its target environment. In contrast, our solution measures relative changes in target performance metrics when available resources are changed. Based on this, trends can be identified to predict the resource requirements in a target environment. Identifying such trends and utilize them is one of the goals of our future work. 
Initial results that proof the general feasibility of our profiling approach are presented in Section~\ref{sec:measurements}.

\subsection{System Design}

\begin{figure}[!ht]
\centering
\begin{subfigure}{.5\columnwidth}
  \centering
  \includegraphics[width=0.8\linewidth]{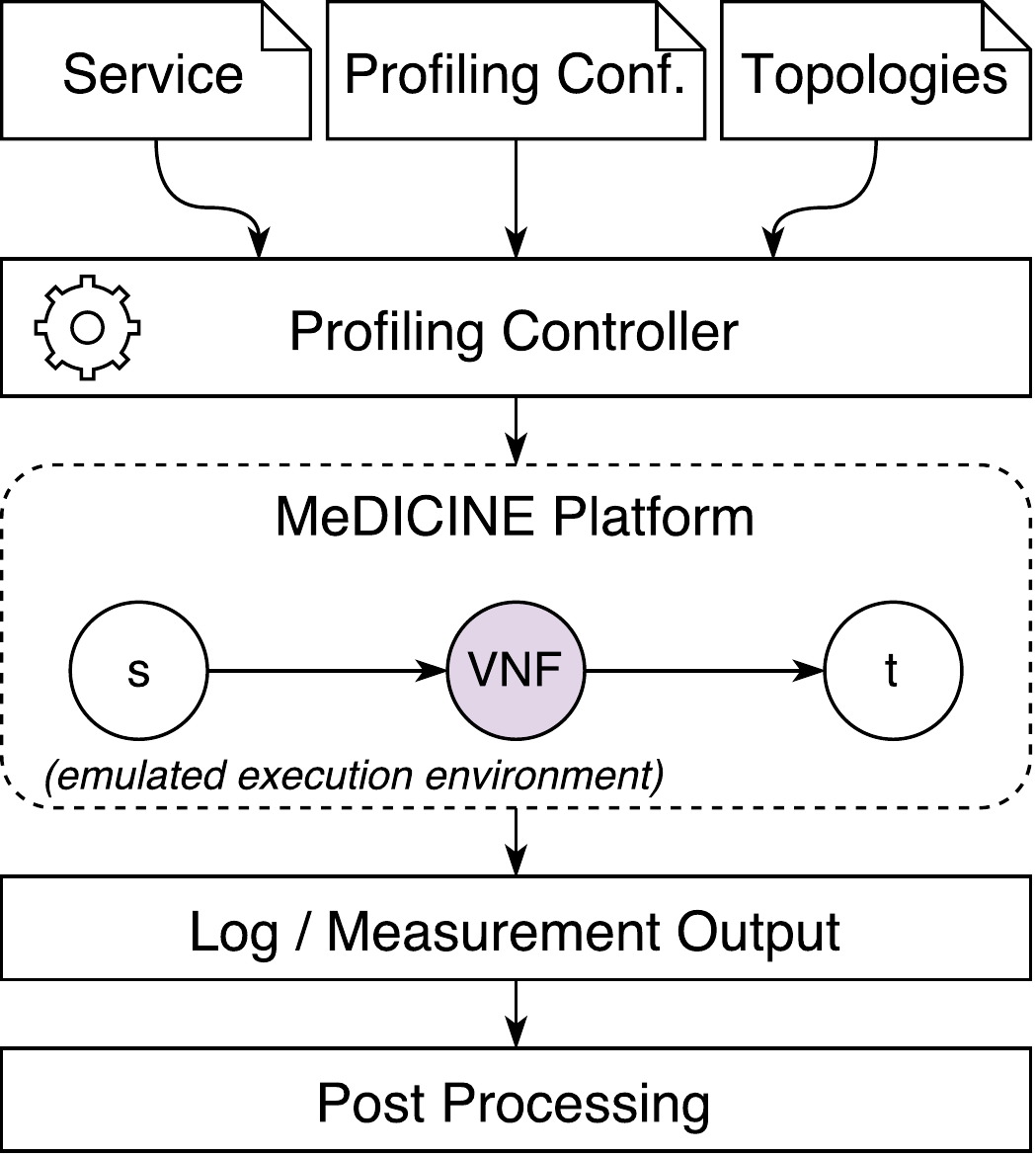}
  \caption{Profiler Prototype}
  \label{fig:system}
\end{subfigure}%
\begin{subfigure}{.5\columnwidth}
  \centering
  \includegraphics[width=1.0\linewidth]{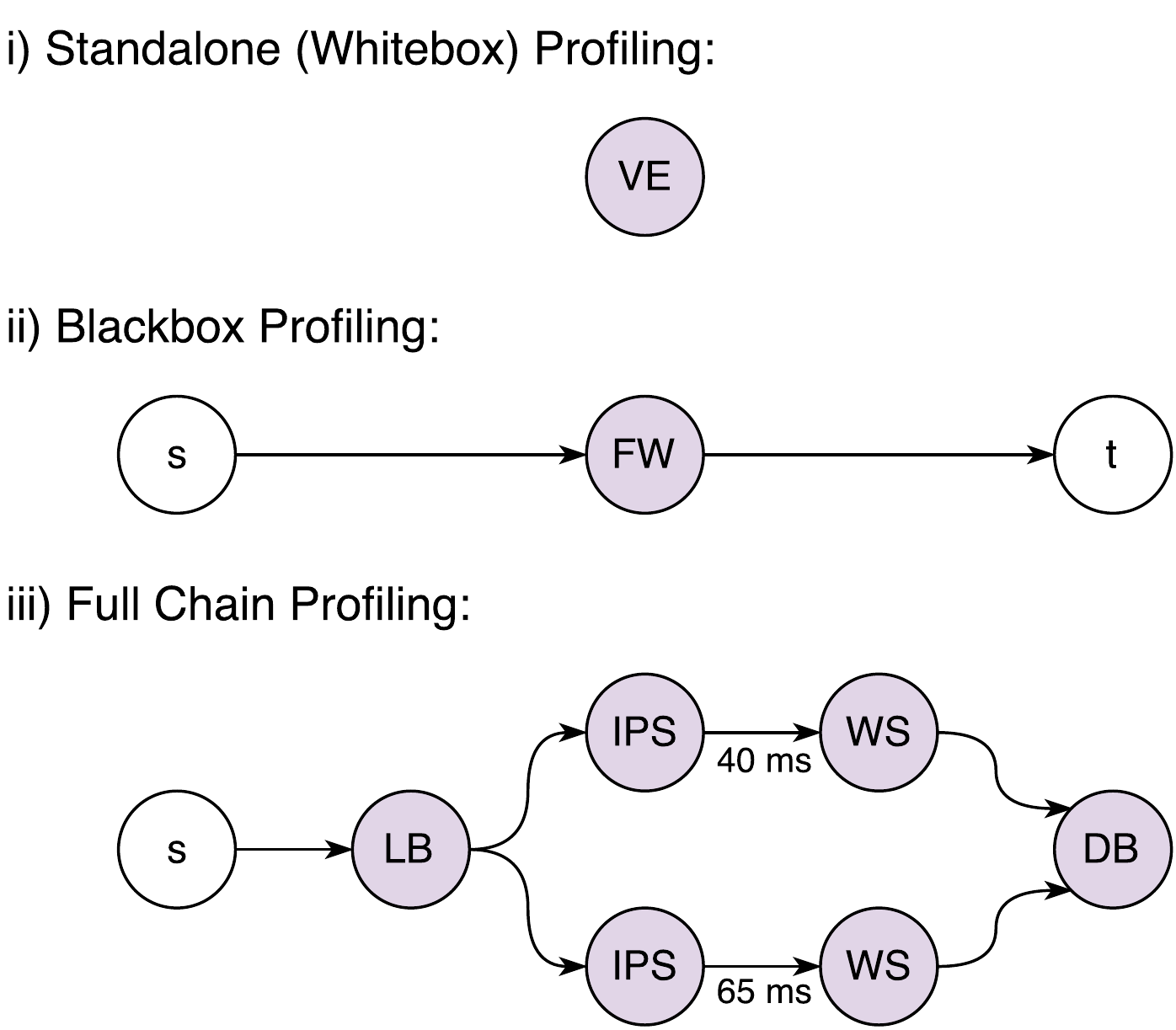}
  \caption{Example Profiling Topologies}
  \label{fig:example}
\end{subfigure}
\caption{(a) Profiler prototype controlling the \emph{MeDICINE} platform to execute network services or VNFs with different resource limitations. (b) Three examples of common profiling topologies supported by our prototype.}
\label{fig:system_total}
\end{figure}

Fig.~\ref{fig:system} shows the design of our prototype. It gets profiling configurations, topology definitions, and container images of the VNFs that should be profiled as its inputs and configures the underlying \emph{MeDICINE} platform accordingly. During a profiling run, the controller iterates over all given resource configurations defined in the profiling configuration and performs the following steps for each of them: Start \emph{MeDICINE} platform, deploy given VNF containers, run additional \emph{probe containers}, e.g., traffic generators, kick-off the measurement process, and collect its outputs. After each cycle, the used containers are terminated and new ones are deployed using another resource configuration.

A single resource configuration can define the following limitations for each profiled container:

\begin{table}[h]
\centering
\begin{tabular}{llp{4cm}} 
\toprule
 \texttt{cpu\_cores} & $\mathbb{N}_{>0}$ & Number of CPU cores \\  
 \texttt{cpu\_time} & \% & Available CPU time \\  
 \texttt{mem\_max} & MByte & Max. available memory \\  
 \texttt{mem\_swap\_max} & MByte & Max. available swap memory \\  
 \texttt{block\_io\_bw} & MByte/s & Max. block device read/write speed \\  
\bottomrule
\end{tabular}
\end{table}

Besides these resource configurations, a developer can also define a topology used for the profiling run. Such topologies can range from very simple ones, containing only a single VNF, to complex ones that represent entire service chains. Fig.~\ref{fig:example} shows three example topologies. The first one shows a single video encoder (VE) function that is profiled without external components by encoding local test files and measuring, e.g., its achieved frame rate under different resource configurations (\emph{whitebox profiling}). The second example shows a so-called \emph{blackbox profiling} scenario (the VNF is not monitored directly) in which traffic is generated at source $s$, sent through the VNF, e.g., a firewall (FW), and received by target $t$. In such a scenario, the resource configuration of the firewall is changed and the throughput is measured at target $t$. 
The additional \emph{probe containers} used for such measurements are either pre-defined or custom tailored by a developer and get isolated resources during a profiling run, e.g., a dedicated CPU core, to not interfere with the profiled VNFs.  
The third example shows a profiling scenario that profiles an entire service chain with multiple functions. Additional network delays are added to the links to emulate the deployment of the chain across multiple PoPs. 
In such scenarios, a resource configuration for a single profiling run must contain resource limits for each of the involved VNFs. For example, a CPU bandwidth configuration for the chain could look like this: \texttt{(LB=0.2;IPS=0.1;WS=0.15;DB=0.05)}. Testing many of these resource configurations helps developers to identify resource relationships and bottlenecks of the chain.

During profiling runs, logging information and performance data is collected by the individual containers and written to shared volumes to be accessible from the host machine. For example, target probe $t$ in Example ii) will store its measurements in such a volume. This approach allows developers to monitor any metric of interest during a profiling run. At the end of this process, post-processing scripts are triggered that collect the stored measurements and process them, e.g., plot them, normalize them, or write them to a database. We do not fix the implementation of these post-processing scripts and allow developers to provide their own to support arbitrary user-defined performance metrics.

\section{Initial Experiments and Results}
\label{sec:measurements}

We conducted a series of initial profiling experiments with different kinds of software applications that are typically part of services executed in a network. We used these experiments to verify the feasibility of our offline profiling approach and to test our prototype. The first set of experiments uses the PTS benchmark suite~\cite{Phoronix.webpage} for whitebox profiling runs (Fig.~\ref{fig:example}). All benchmarks have been performed with the default configurations of the PTS suite and have been repeated three to five times according to these default configurations. The benchmarking suite was installed in an Ubuntu 14.04-based Docker container and the experiments have been executed on a machine with Intel(R) Core(TM) i7-960 CPU @ 3.20~GHz, 4 physical cores, hyper threading, and 24~GB RAM. The error bars in all presented results represent 95\% confidence intervals.

\begin{figure}[!ht]
\centering
\begin{subfigure}{.3\columnwidth}
  \centering
  \includegraphics[width=1.0\linewidth]{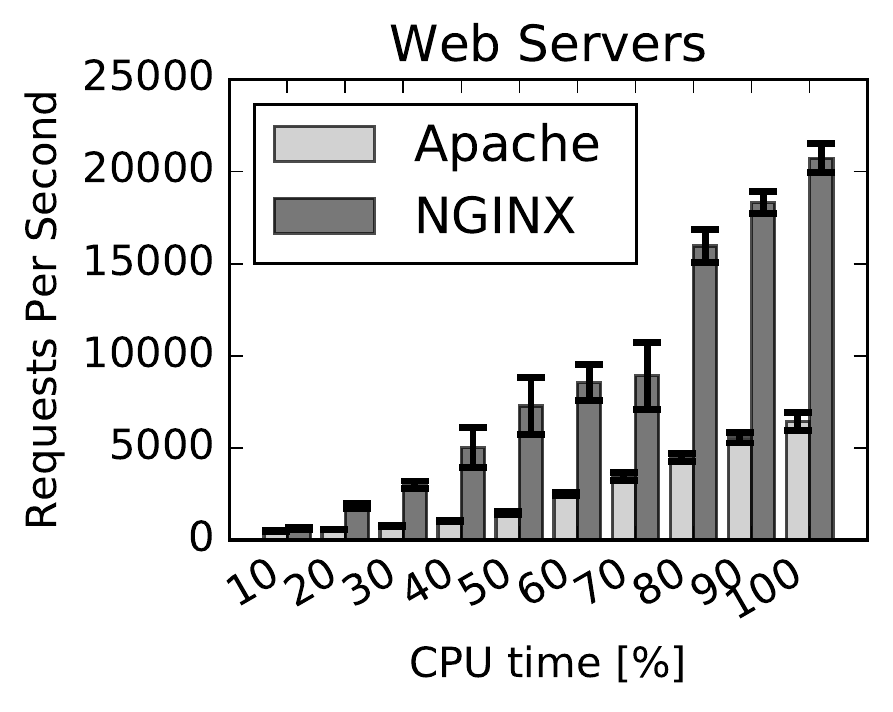}
  \label{fig:plot11}
\end{subfigure}%
\begin{subfigure}{.3\columnwidth}
  \centering
  \includegraphics[width=1.0\linewidth]{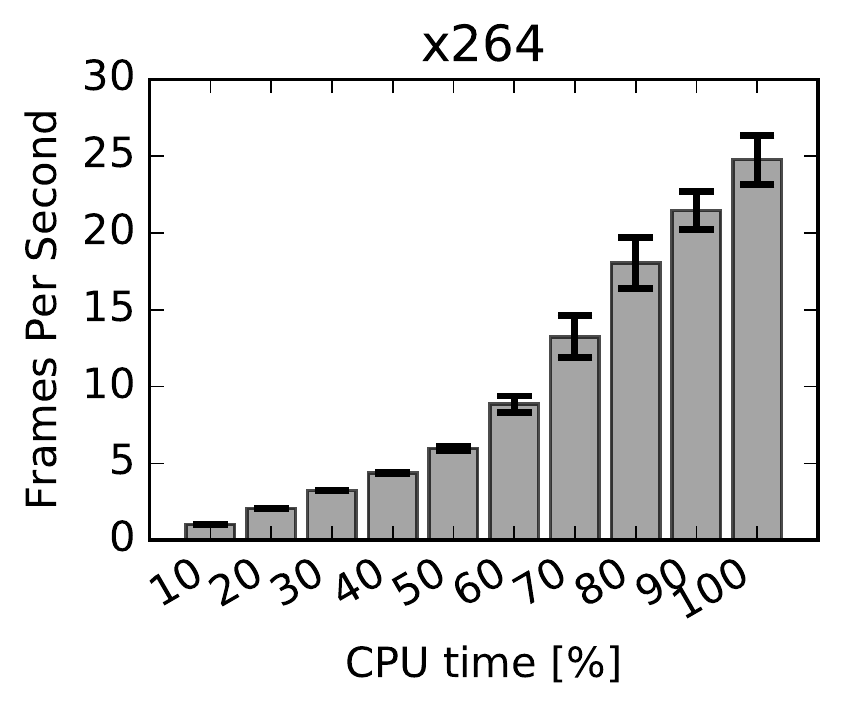}
  \label{fig:plot12}
\end{subfigure}%
\begin{subfigure}{.3\columnwidth}
  \centering
  \includegraphics[width=1.0\linewidth]{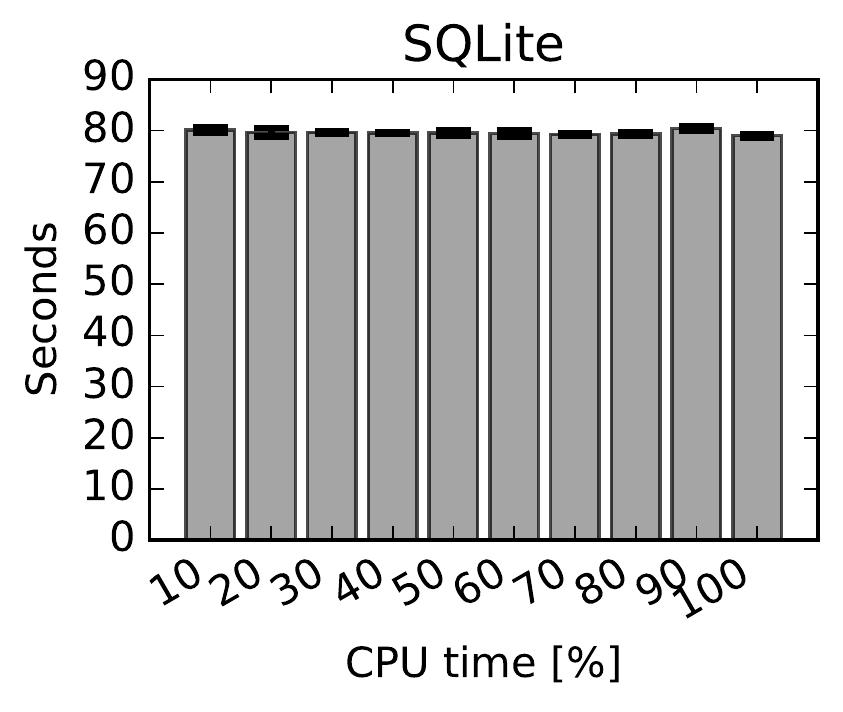}
  \label{fig:plot13}
\end{subfigure}

\begin{subfigure}{.3\columnwidth}
  \centering
  \includegraphics[width=1.0\linewidth]{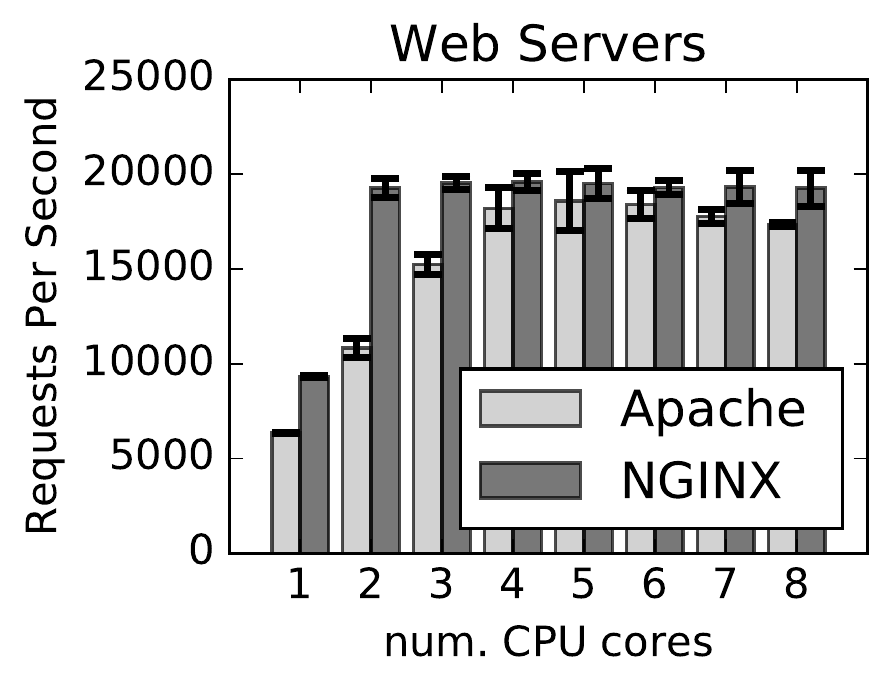}
  \caption{Web servers}
  \label{fig:plot14}
\end{subfigure}%
\begin{subfigure}{.3\columnwidth}
  \centering
  \includegraphics[width=1.0\linewidth]{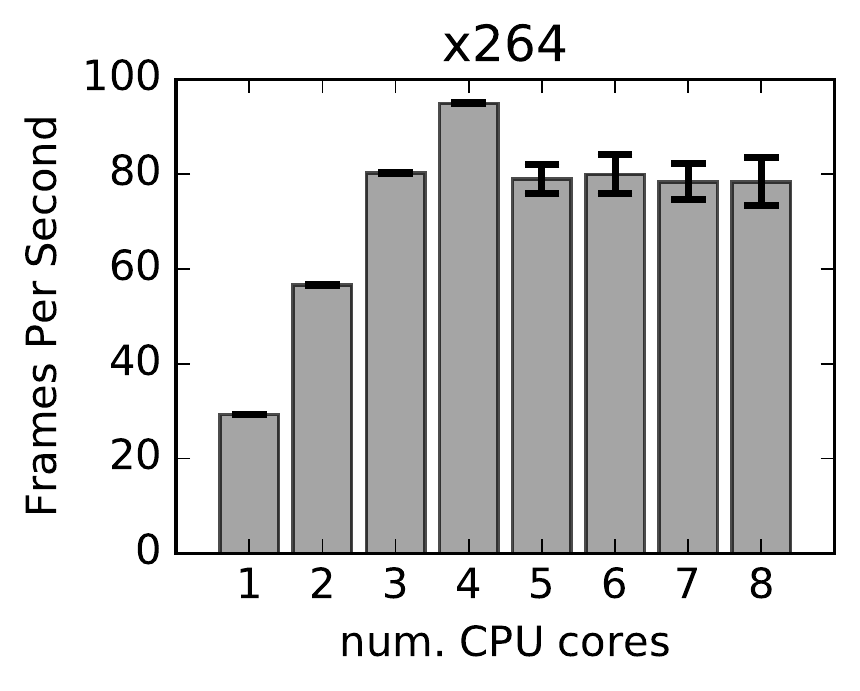}
  \caption{Video encoding}
  \label{fig:plot15}
\end{subfigure}%
\begin{subfigure}{.3\columnwidth}
  \centering
  \includegraphics[width=1.0\linewidth]{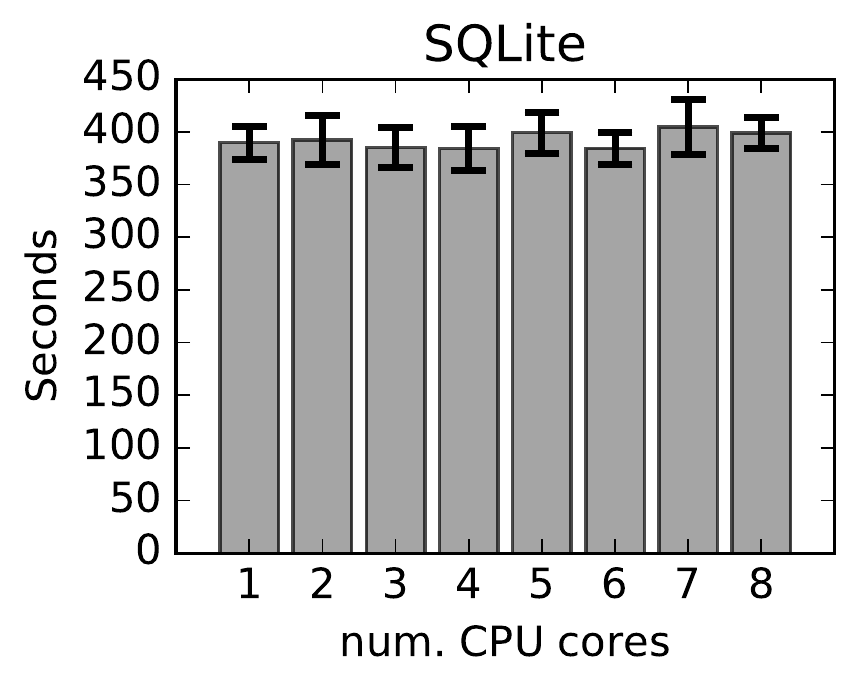}
  \caption{Database}
  \label{fig:plot16}
\end{subfigure}

\caption{Application performance under different CPU configurations}
\label{fig:plot1}
\end{figure}

Fig.~\ref{fig:plot1} shows results for different CPU resource configurations. In the first three experiments, we allocated different amounts of CPU time to the profiled containers and let them use all available CPU cores. In the second set of experiments, we allocated different numbers of CPU cores to the containers. The results show that the impact of available CPU resources to an application's performance depends on the type of application. As expected, a database application is not bound by CPU resources. This draws the case for offline profiling solutions because a MANO system needs this kind of information to improve its scaling decisions, e.g., adding additional vCPUs to a database VNF will not result in performance improvements.  
In addition, the results indicate that even applications of the same type can have different scaling behaviors. For example, the Nginx web server already achieves its maximum performance when the second CPU core is added, whereas the Apache web server shows a linear pattern when the number of cores is increased. The figure also shows that there are no performance improvements after the maximum of four physical cores is allocated. Especially Fig.~\ref{fig:plot15} shows this effect that occurs when more than the four available physical cores are assigned using the hyper threading functionality of the CPU. 
The experiments demonstrate that our profiling approach is able to measure different performance metrics including application-level metrics, like \emph{frames/s} in the \emph{x264} video encoder experiment. 

The second set of experiments represents the setup of the second example shown in Fig.~\ref{fig:example}. These experiments use iperf-based probe containers to send traffic through a \emph{snort}~\cite{Snort.webpage} intrusion detection VNF and measure its throughput. 
We profiled two major versions of \emph{snort}, namely \emph{version~2.9} installed from Ubuntu's package repositories and \emph{version~3.0alpha} which is a preliminary release available as source code. Both versions are used in their default configuration and we profiled them under changing CPU limitations.

\begin{figure}[!ht]
\centering
\begin{subfigure}{.4\columnwidth}
  \centering
  \includegraphics[width=1.0\linewidth]{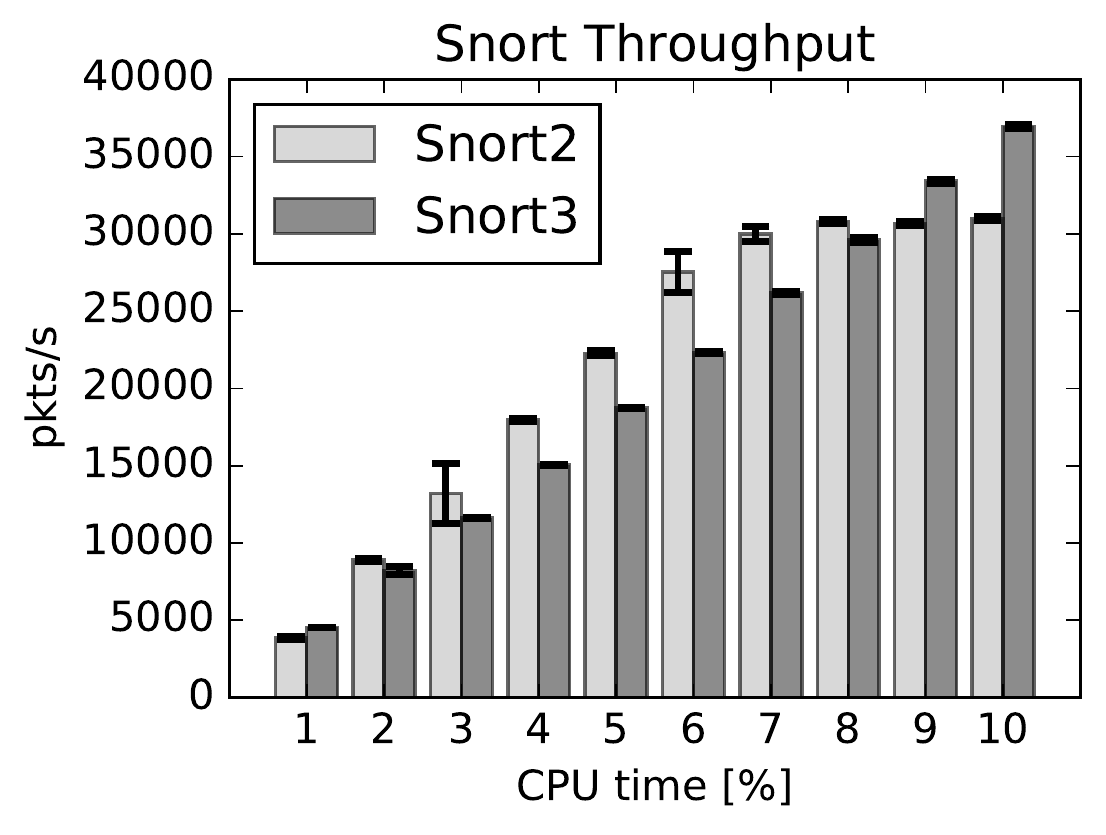}
  \caption{Limited CPU time}
  \label{fig:plot21}
\end{subfigure}%
\hspace{.5cm}
\begin{subfigure}{.4\columnwidth}
  \centering
  \includegraphics[width=1.0\linewidth]{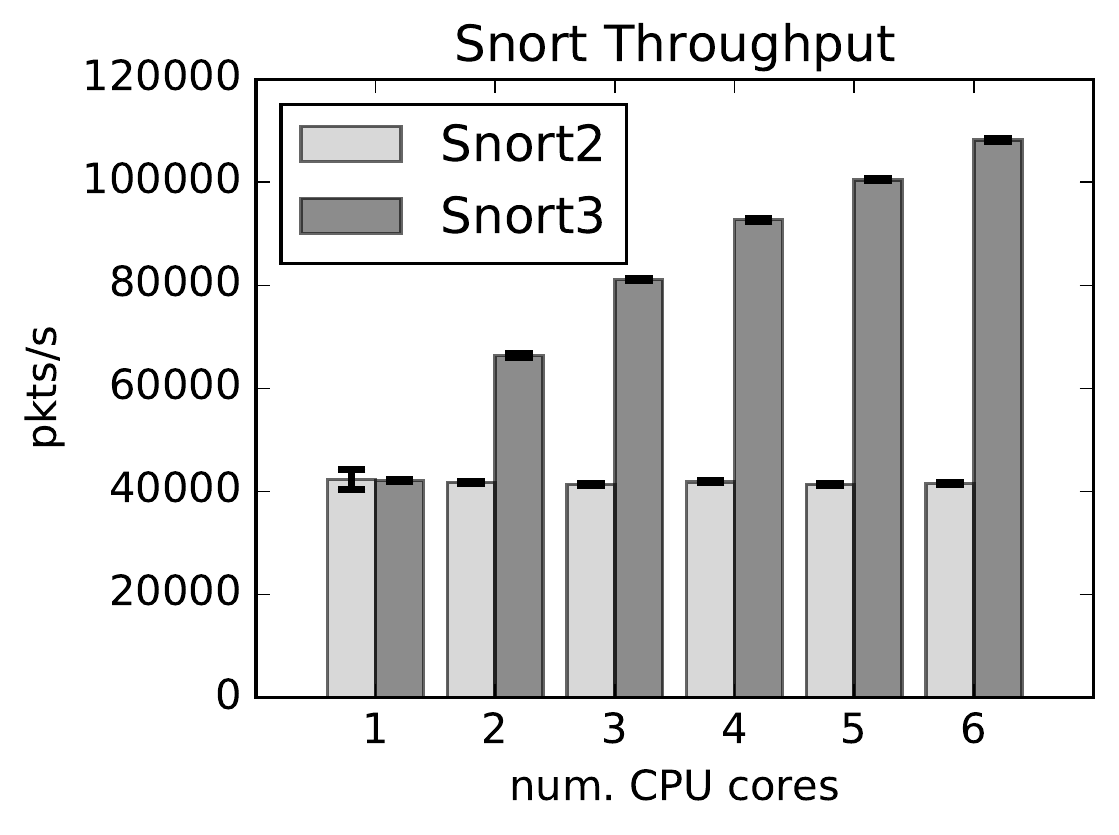}
  \caption{Number of CPU cores}
  \label{fig:plot22}
\end{subfigure}%

\caption{Comparison of two major versions of the \emph{snort} IDS system under changing CPU configurations}
\label{fig:plot2}
\end{figure}

Fig.~\ref{fig:plot2} shows the averaged results of 25 repetitions of these experiments. The results provide insights about the runtime behavior of these two different \emph{snort} versions. Fig.~\ref{fig:plot21} shows their behavior under very limited CPU time allocations ($\leq 10\%$) on a single core under which both \emph{snort} versions behave almost identical. In contrast to this, does \emph{version~3.0alpha}  outperform the old \emph{snort} version when the number of available CPU cores is increased (Fig.~\ref{fig:plot22}). The obvious reason for this is the fact that the new \emph{snort} version introduces multithreading. This example clearly shows that the behavior of a VNF can drastically change between versions and that the generated profiling results can support a MANO system to automatically adapt its resource allocation decisions to these changes.

\section{Related Work}
\label{sec:rw}

Applying profiling solutions to improve NFV resource management is still a novel research direction with limited related work. However, there was a lot of work done by the cloud computing community that includes profiling mechanisms as well as forecasting and prediction approaches~\cite{weingartner2015cloud}. Several of these solutions rely on dedicated cloud testbeds to profile applications. This makes them hard to apply to an agile DevOps methodology~\cite{do2011}\cite{giannakopoulos2015panic}. Other solutions execute applications directly on physical machines to measure their performance offline and try to predict the application's performance in virtualized environments~\cite{tak2013pseudoapp}\cite{wood2008profiling}. They do not provide fine-grained control about resource constraints under which an application is executed, e.g., number of CPU cores. This limits the expressiveness of the resulting profiles compared to our solution. 

Most existing approaches allow profiling of single applications~\cite{do2011}\cite{wood2008profiling}\cite{dawoud2012dynamic} and only some of them can profile complex, composed applications~\cite{giannakopoulos2015panic}\cite{tak2013pseudoapp}. However, none of them focuses on NFV service chains nor do they allow to configure different resource limits for subparts of the profiled application which is supported by our approach. 


The most relevant work is presented in \cite{rosa2015vbaas} and proposes a solution called VNF benchmarking as a service (VBaaS). It  is a framework to construct NFV infrastructure (NFVI) and VNF performance profiles while avoiding overheads for continuous monitoring. The goal of VBaaS is aligned with the goals of our work that is to endorse the resource allocation decisions made by MANO systems. It provides two kinds of profiles. First, NFVI profiles that describe the performance of certain test VNFs in a given NFVI. Second, VNF profiles that describe the VNF resource consumption for given performance requirements. These profiles are generated by deploying test VNFs, monitoring agents, or probes inside the actual NFVI infrastructure. Thus, their approach does not provide the possibility to do offline profiling  nor is it possible to do specialized profiling runs of entire service chains in custom topologies.

Other work that motivates the need for NFV profiling solutions can be found in an IETF internet draft that describes high-level requirements to benchmark virtualized network functions and services~\cite{rfc.draft}.

\section{Conclusion and Future Work}
\label{sec:conclusion}

In this position paper we discussed and motivated the need to integrate profiling into the NFV development cycle. We outlined a first architectural concept that indicates that several new components and artifacts, such as performance profiles for VNFs and network service chains, have to be added to existing NFV workflows and MANO systems. Consequently, we presented an early prototype of an offline profiling system that allows to profile VNFs and entire services under realistic resource constraints on a local machine. The key insight obtained from first experiments done with our prototype is that even within the same class of applications the performance behavior under different resource constraints differs and that such information has to be made available to MANO systems to improve decision precesses.

We plan to continue our work in this field by following the research questions outlined in Section~\ref{sec:questions}. We will especially focus on finding automated solutions to normalize profiling results and on designing prediction systems that utilize profiling information to improve resource allocation decisions, e.g., by using machine learning approaches. Further, we will continue our work on our offline profiling prototype and plan to open source the tool as part of our \emph{MeDICINIE} emulation platform.

\section*{Acknowledgments}
\footnotesize{This is a pre-print version. The original paper was submitted to and accepted by the European
  Workshop on Software Defined Networks (EWSDN) 2016. This work has been partially supported by the SONATA project, funded by the European Commission under Grant number 671517 through the Horizon 2020 and 5G-PPP programs (\url{www.sonata-nfv.eu}) and the German Research Foundation (DFG) within the Collaborative Research Centre ``On-The-Fly Computing" (SFB 901).}

\bibliographystyle{IEEEtran}
\bibliography{IEEEabrv,main}

\end{document}